\input{aipcheck.tex}

  \def\selectedoptions{final}

\documentclass[
   \selectedoptions
  ]
  {aipproc}

\layoutstyle{6x9}

\SetInternalRegister\hbadness{8000} 

\newcommand\doingARLO[2][]{%
  \ifx\mmref\undefined #1\else #2\fi
}
  
\def\approxgt{\,\raise2pt \hbox{$>$}\kern-8pt\lower2.pt\hbox{$\sim$}\,}
\def\approxlt{\,\raise2pt \hbox{$<$}\kern-8pt\lower2.pt\hbox{$\sim$}\,}
\def\th{\thinspace}

\begin{document}

\title 
      [Chaos in the Music of the Spheres]
      {Chaos in the Music of the Spheres}

\classification{43.35.Ei, 78.60.Mq}
\keywords{}

\author{J. Robert Buchler}{
  address={University of Florida, USA},
  email={buchler@phys.ufl.edu},
  thanks={Work supported by the NSF}
}

\iftrue
\author{Zoltan Koll\'ath}{
  address={Konkoly Observatory, HUNGARY},
  email={arno@mittelbach-online.de},
  thanks={Work supported by MTA-OTKA (T-026031)}
}

\author{Robert Cadmus}{
  address={Grinnell College, USA},
  email={cadmus@grinnell.edu},
  thanks={Work supported by the NSF}
}
\fi

\copyrightyear  {2001}

\begin{abstract} The light curves (time series of the radiated energy) of most
large amplitude, pulsating stars such as the well known Cepheid stars are
regular.  However, a smaller group of variable stars that are located next to
them in the Hertzsprung- Russell diagram undergoes irregular light variations
and exhibits irregular radial velocities as well.  The mechanism behind this
irregular behavior was a long standing mystery.  A flow reconstruction
technique based on the observed lightcurves of six separate stars shows that
their underlying dynamics is chaotic and low dimensional ($d=4$).  Furthermore,
we present evidence that the physical mechanism behind the behavior is the
nonlinear interaction of just two pulsation eigenmodes.  In a generalized
Shil'nikov scenario, the pulsation energy alternates continuously, but
irregularly between a lower frequency mode that is linearly unstable and thus
growing, and a stable overtone that gets entrained through a low order
resonance (2:1), but that wants to decay.  The flow reconstruction from the
stellar light curve thus yields interesting physical insight into the pulsation
mechanism.  \end{abstract}

\date{\today}

\maketitle

\section{Introduction}

Almost every type of star becomes unstable to self-excited vibrations at least
once during its lifetime.  Typically, the stars that exhibit large amplitude
variations in light are undergoing radial pulsations, {\it i.e.} pulsations in
which the star remains spherically symmetric at all times.  Typically, only the
fundamental pulsation mode or the first two overtones are involved in the
pulsation.  In contrast, most small amplitude variable stars such as the Sun or
white dwarfs undergo nonradial pulsations, but with many modes participating in
the motion.

The classical Cepheid variables and RR Lyrae variables are the best known and
best studied variable stars of the radial pulsator category.  They owe this
interest at least in part to their role as primary cosmological distance
indicators; they are almost perfectly periodic and obey a period-luminosity
relation that can be used for measuring their distances.

This paper concerns itself with variable stars that lie next to and slightly
below the classical Cepheids in a long and broad curved strip in the
Hertzsprung - Russell diagram (luminosity vs. temperature diagram) \cite{cox}.
For historical reasons the stars at the lower luminosity end of the strip are
called W~Virginis type stars, further up RV~Tauri type stars, then Semiregular
stars and finally Mira variables, even though the transition between the
different classes appears to be gradual.  For convenience we shall refer to all
of them here as semiregular stars, {\it largo sensu}.  The W~Vir stars that lie
at the low luminosity end of the strip with concomitantly low period
($P\approxlt 20d$) have regular light curves, but for the slightly more
luminous stars stars the observations show alternations in the pulsation cycles
that appear to be due to period doubling \cite{arp55}.  In fact, numerical
hydrodynamical models display cascades of period doublings \cite{b87},
\cite{pdchaos88}, \cite{aik90} as well as tangent bifurcations \cite{bgk87},
\cite{aik87}.  Observations also show that stars with higher luminosity have
increasingly irregular pulsations \cite{pol96}.

\begin{figure}
  \includegraphics[height=0.8\textheight]{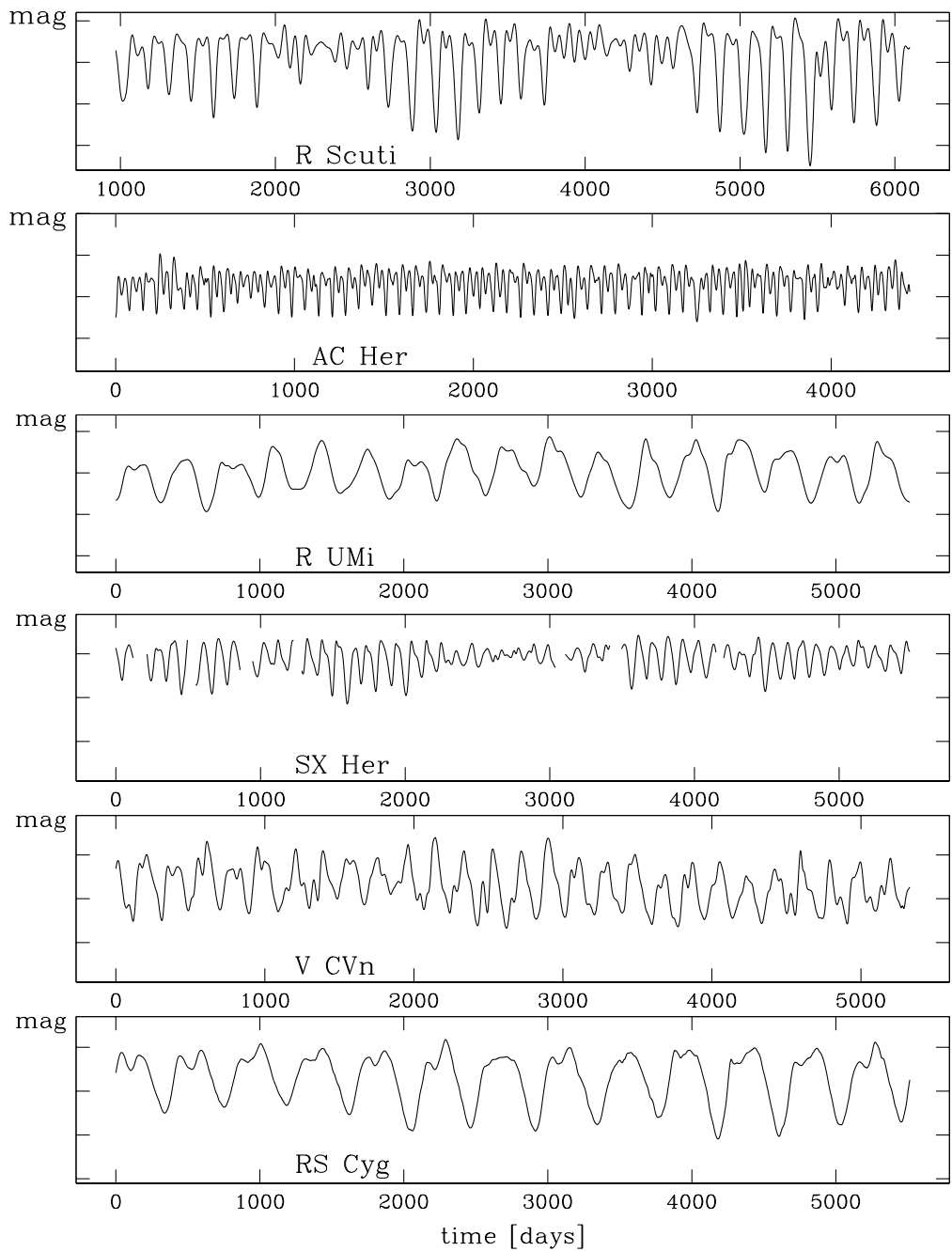}
  \caption{Smoothed observed light curves}
   \label{figLC}
\end{figure}

Traditional astronomers like to think of stars are being multi-periodic
(meaning that they consider the pulsation to be a superposition of a
number of pulsation frequencies with steady amplitudes).  It is true that a
phenomenologically motivated multiperiodic fit will always be 'successful' (as
an interpolation), but on physical grounds it is possible to rule out such a
description \cite{takeuti}.  In a nutshell:

(1) The frequency peaks in the Fourier spectrum of the light curve are
numerous, but in this type of star, and in the frequency range of interest,
there do not exist enough radial pulsation eigenmodes, nor even low order
nonradial pulsation modes that could be observable and excitable to such large
amplitudes.  No multi-periodic fits are possible that can reproduce the light
curves within the observational accuracy while using modal frequencies that are
acceptable on a {\sl physical basis}.

(2) The frequencies in the Fourier spectra appear randomly variable from one
section of the light curve data to another (for R~Sct {\it cf}.
\cite{kollath90}).  If a multi-periodic star were evolving slowly, the
structure of the peaks should evolve slowly as well.  The absence of such
correlations eliminates any explanation of the lightcurve as that of slowly
evolving multi-periodic stars.

(3) One recent paper (\cite{konig99}) 'explains' the irregular behavior of
R~Sct as the superposition of stochastically excited linear oscillators.  While
such a stochastic 'explanation' is perhaps mathematically valid, it is not
meaningful on physical grounds because no mechanism is proposed (nor can be
found) that could excite damped modes to such large pulsation amplitudes (up to
factors of 40 in the light curve of R~Sct).

\vskip 2mm

Unfortunately the cycling times of these variable stars vary from as low as 20
days for the lower luminosity W~Vir stars to as much as a year for the
Semiregular variables.  It is almost needless to say that very few stars have
been observed with a coverage that lends itself to modern nonlinear analyses
such as a flow reconstruction.  Furthermore, for obvious reasons, the
observations are generally not spaced with equal time intervals, so that an
interpolation with concomitant introduction of noise is unavoidable.

\begin{figure}
  \includegraphics[height=0.8\textheight]{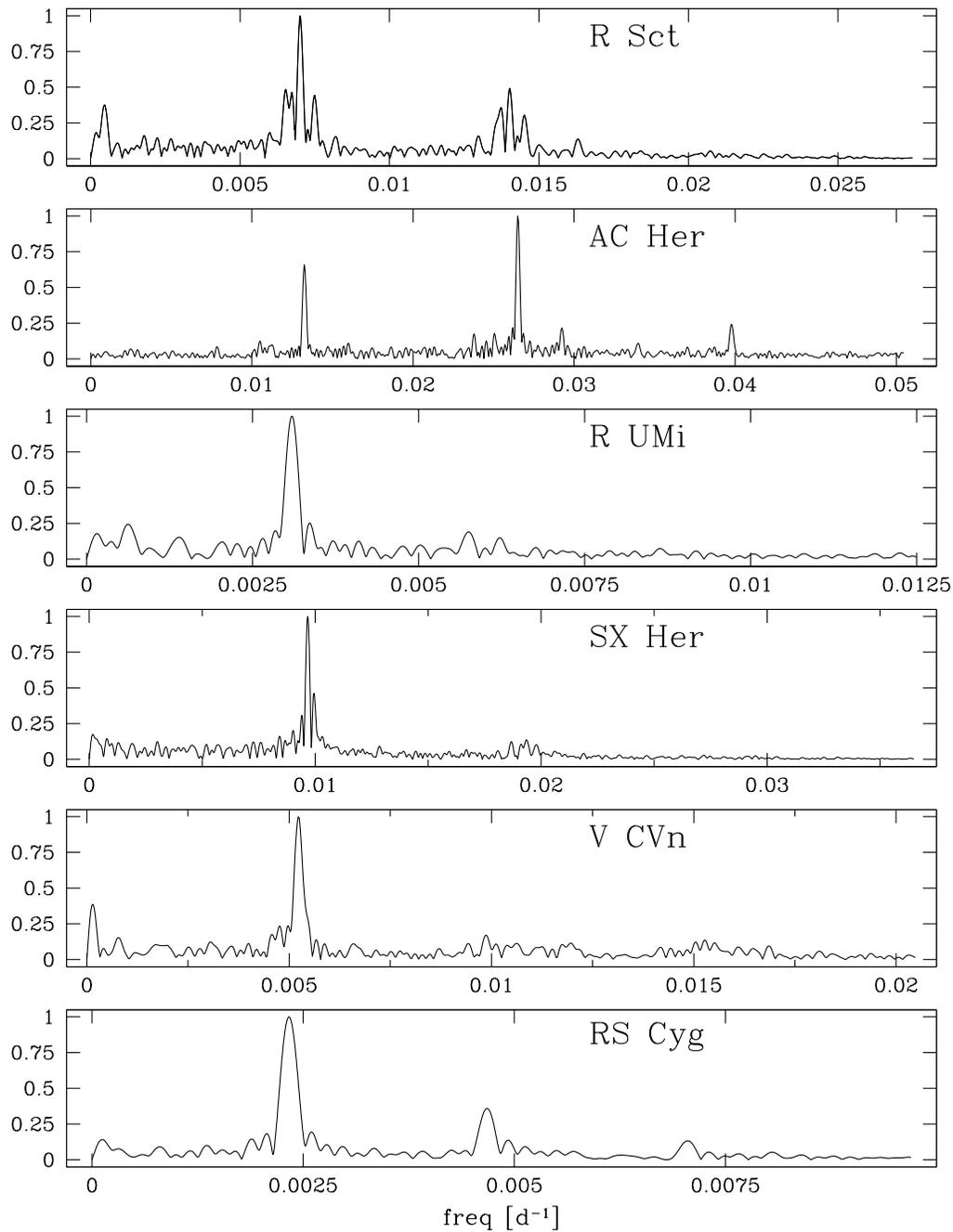}
  \caption{Amplitude Fourier spectra (relative amplitude scale)}
   \label{figFS}
\end{figure}

A few years ago we were fortunate to obtain access to and analyze the data sets
of two irregular variable stars, called R~Scuti \cite{rsctprl95} \cite{rsct96}
and AC~Herculis \cite{acher98}, that have a sufficiently good
coverage and exhibit a large enough number of pulsation cycles
(cf. Fig.~\ref{figLC}) to be representative of their long term behavior.

Our analysis was based on a flow reconstruction with a multivariate polynomial
map ${\cal M}$ (or o.d.e.) in an embedding space of dimension $d_e$
$$X_{n+1} = {\cal M} \cdot X_n$$
\noindent where the state vector $X\in {\bf\rm R}^{d_e}$ 
is constructed from the scalar time-series
$\{x_i\}$,
$$X_n = (x_i, \thinspace x_{i-\tau}, \thinspace  x_{i-2 \tau}, \thinspace  
\ldots  \thinspace  x_{i-(d_e-1) \tau} ) $$
\noindent where $\tau$ is a delay parameter that should be long enough so that
noise does not kill the reconstruction, but small enough so that the map does
not become too nonlinear \cite{serre96}.
We briefly summarize here the results that were obtained.

\subsection{The stars R~Scuti and AC~Herculis}

The light curve data of R~Sct were shown to be generated by a 4-dimensional
dynamics (d=4) \cite{rsctprl95} \cite{rsct96}.  This conclusion was based on
several facts: 

\noindent \ (a) the minimum embedding dimension is $d_e=4$: first, synthetic
light curves generated in 3D bear no resemblance to the observations; second,
the error of the fit levels off at $d_e=4$; third, the nearest neighbor method
indicates 4 as the minimum dimension.  However, the strongest argument comes
from a comparison of the Fourier spectra of the synthetic light curves
(generated through an iteration of the map) with those of that observational
data, and of a comparison of the respective Broomhead-King projections (onto
the eigenvectors of the correlation matrix).

\noindent \ (b) the fractal (Lyapunov) dimensions \cite{ott} derived from the
Lyapunov exponents of the synthetic signals fell in the range 3.1 -- 3.2 for
R~Sct, {\sl independently of the embedding dimension}.  This is therefore a
lucky situation where the bounds $d_L \sim 3.15 < d \leq d_e=4$ uniquely
determine the physical dimension to be $d=4$.

\noindent \ (c) There is no {\it a priori} guarantee that a map or flow should
capture the underlying dynamics.  The fact that one of the Lyapunov exponents
is always close to zero very strongly suggests that we have been successful in
the flow reconstruction. Indeed, a map with a short time-step should be close
to a flow for which we know that the corresponding Lyapunov exponent is exactly
zero.  Actually we have also reconstructed a true flow (system of o.d.e.'s) for
R~Sct, although we find that the reconstruction of the flow is a little less
robust than that of the map.

\noindent \ (d) Perhaps the most interesting physical result comes from the
linearization of the successful maps around their fixed points: Two spiral roots
$\pm i\omega + \xi$ are found with the following properties: $\omega_2 \approx
2 \omega_1$, with $\xi_2 > 0$, and $\xi_1 < 0$, and $|\xi_2| >\xi_1$.  This is
of course reminiscent of the Shil'nikov criterion \cite{tresser}.

These results allow us to give a physical interpretation of the motion.  We
already mentioned that the regular, classical Cepheid variables undergo
pulsations in one of the two lowest modes of pulsation.  It is seen that the
semiregular stars to the same: the complex amplitudes of vibrational modes are
the 'natural' coordinates for the phase space, so that $d=4$ implicates the
involvement of two vibrational modes.  Furthermore, the low frequency mode, of
frequency $\omega_1$ is linearly unstable (self-excited) and grows in
amplitude.  In the nonlinear regime it interacts with a second mode because of
a low order resonance condition ($\omega_2= 2 \omega_1 $).  However, this
entrained mode is linearly stable and wants to decay.  A chaotic motion of
alternating growth and decay ensues.

\vskip 2mm

Our analysis of AC~Herculis \cite{acher98} was a little less conclusive.  The
constructed maps and flows were less robust.  By that we mean that for many
seed values the iteration blew up much more rapidly.  We believe that this lack
of robustness may be due to the lower signal to noise ratio (Fig.~\ref{figLC})
for AC~Her.  Despite these short-comings of the reconstruction, however,
everything points at the minimal embedding dimension again being 4.  The
fractal dimension $d_L$ turned out to be lower $d_L \sim 2.3$ than for R~Sct.

As Fig.~\ref{figLC} shows the light curve of AC~Her has a lot smaller amplitude
swings, and hardly any phases of low amplitude oscillations.  The linear part
of the map which describes the vicinity of the fixed point is therefore less
well determined, and so are the stability roots.  We were therefore unable to
verify whether the same resonant entrainment as in R~Sct is operative in
AC~Her.

\section{New results}

In this paper we present an analysis of observational data of four additional
stars, taken by one of us (RC) at Grinnell College over a 15 year period
(cf. Fig.~\ref{figLC}).  These data are much higher quality than the amateur
astronomer data previously analyzed, but they suffer from the drawback of
lesser coverage.  The stars go under the names of SX~Herculis, R~Ursae Minoris,
RS~Cygni and V~Canum Venaticorum, and astronomers classify them as Semiregular
variables (or Mira type for V~CVn).  The light curves of these stars are shown
together with those of R~Sct and AC~Her which we analyzed earlier.  The light
curves of Fig.~\ref{figLC} represent cubic spline fits to the observational
data.

{\begin{figure}
  \includegraphics[height=0.38\textheight]{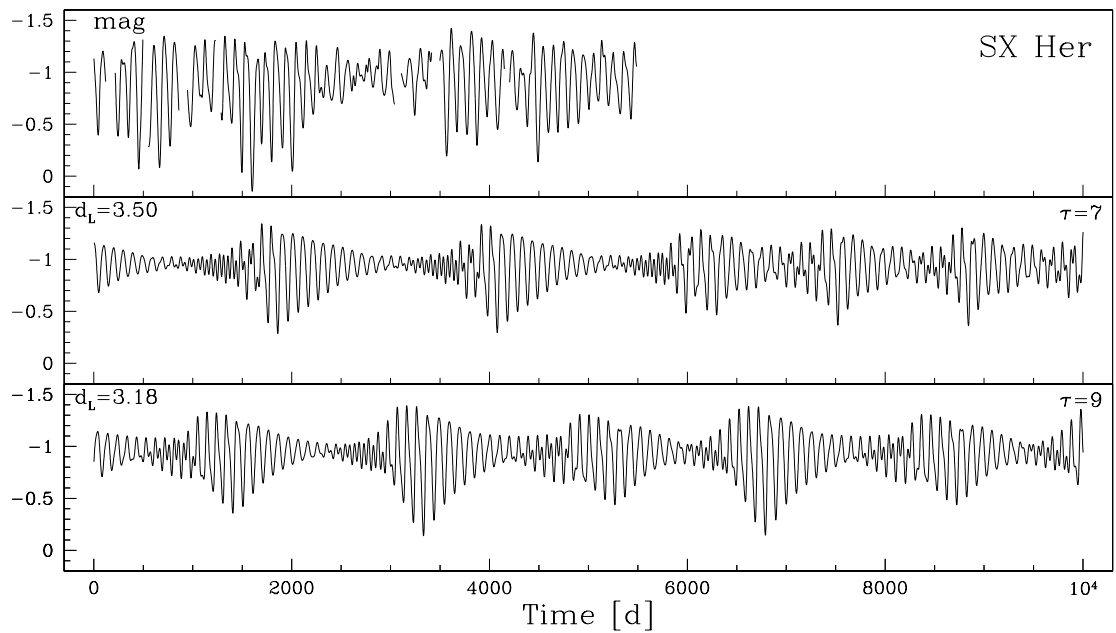}
  \caption{Flow reconstructions with the light curve magnitude of SX~Her.}
   \label{sxhers}
\end{figure}
\begin{figure}
  \includegraphics[height=0.38\textheight]{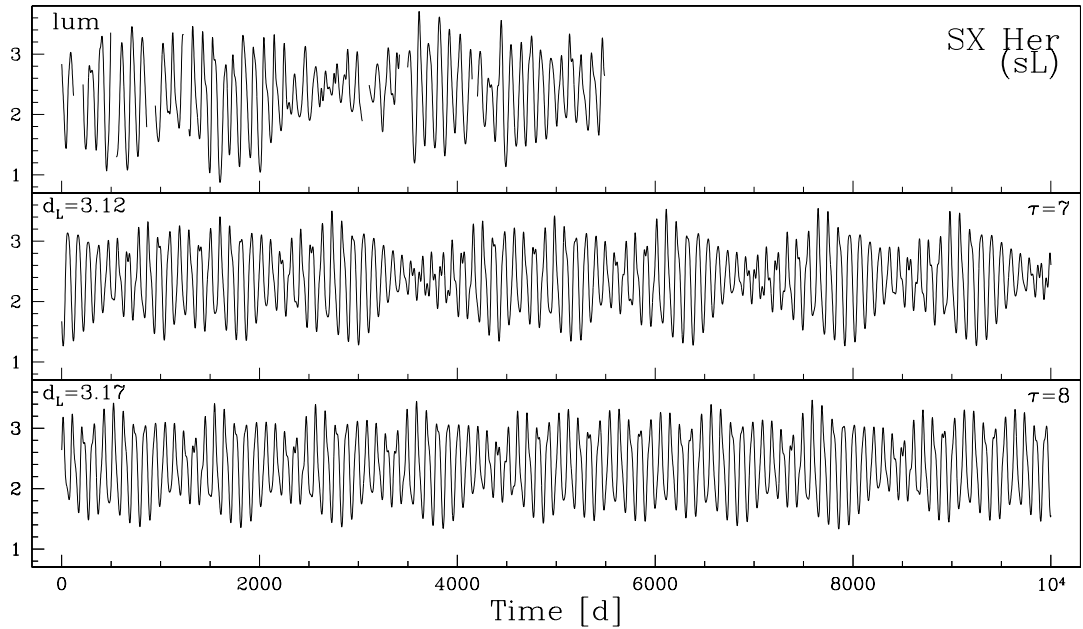}
  \caption{Flow reconstructions with the luminosity of SX~Her.}
   \label{sxhersL}
\end{figure}
}

Our cubic spline fits depend on the amount of smoothing which is controlled by
the parameter $\sigma$ (\cite{reinsch}), and by the size of the gaps that we
allow.  Interpolation is necessary because our analysis requires a time series
with equal time intervals.  Typically we use a value of $\sigma$ = 0.02.

In Fig.~\ref{figFS} we also display the amplitude Fourier spectra of the six
stars.  Of the four stars RS~Cyg has the most harmonic power whereas SX~Her has
almost none.  But, as is well known, we do not learn too much about the
dynamics from linear analyses, such as Fourier, MEM, ARMA or time-frequency
\cite{weigend}.

\subsection{The star SX Herculis}

The smoothed 5500\th day long observational light curve (with 1~day sampling)
of SX~Her is shown on top of Fig.~\ref{sxhers} for reference.  In the flow
reconstruction we have used astronomical magnitudes, $m \equiv - 2.5 {\rm Log}
L$, rather than the more physical luminosity ($L$, energy radiated per unit
time).  Underneath, we display two of the best results of our global flow
reconstructions in the form of segments of synthetic light curves that were
generated with the help of 4D maps and two different values of $\tau$,
indicated in the upper right corner of the graphs.  The synthetic light curves
have been obtained through a 10500 fold iteration of the maps with different
initial seeds, thus generating 10000\th d long signals.  (We have discarded the
first 500 iterations to avoid transients).

We note however that for a given map not all seeds lead to this type of signal,
some go to fixed points, some blow up.  We attribute these failures at least
partially to the shortness of the observational data set that was used to train
the map.  After all, the observational signal samples only a small part of
phase space, and not very densely for that matter.  As the map is iterated the
trajectory eventually ends up in a part of phase space where the map is not
well determined because of the training signal had no or few points there, and
it runs off to infinity.  For other parameter values the opposite can also
occur, namely that the map is too stable leading to limit cycles because of a
paucity of phase space points around this limit cycle.  Stable synthetic
signals resembling the observations could only be found for some values of
$\tau$ (7 and 9), indicating that this reconstruction lacks robustness.
Despite these shortcomings, however, the synthetic light curves display many of
the characteristic features of the observed light curves, such as asymmetric
bursts, but they are somewhat more regular.

In the upper left corners we show the Lyapunov dimensions of the synthetic
signal.  They have been computed with relatively short time-series (10000
points) and are thus not very precise.  Nevertheless the maps that
successfully produce synthetic signals with the characteristics of the
observations all have a fractal dimension $d_L$ in the range from 3.1 -- 3.7.

Reconstructions in 3D are not successful at all.  In 5D they are successful,
but not very robust, most likely because of the short number of modulation
cycles.  However, the few synthetic signals that we were able to reconstruct
had a fractal dimension less than 4, despite the fact that the embedding space
was 5.  This is of importance because it suggests again, as for R~Sct that the
physical dimension is $d=4$.

When we linearize our 4D maps around their fixed point, we find that for the
good reconstructions the fixed points of the maps are of a doubly spiral
nature, as in the case of R~Sct. (There can be more than one fixed point, but
generally only one is located at the 'center' of the motion).  The ratio of the
two frequencies is found to be 2.1 -- 2.2, and the stable growth rate of the
higher frequency root (real part of the linear spiral eigenvalues) is always
larger than the unstable one, again a generalized Shil'nikov scenario for chaos
\cite{tresser}.  The linearization of the map is of course is closely related
to the study of the linear vibrational properties of the equilibrium star.
Thus for SX~Her, as for R~Sct earlier on, we arrive at a physical picture of
the irregular pulsation mechanism, namely one of continual and alternating
exchange of energy between a growing pulsation mode of frequency $f_0$ and a
decaying resonant mode of frequency $\sim 2f_0$.

\vskip 0.2cm

\begin{figure}
  \includegraphics[height=0.38\textheight]{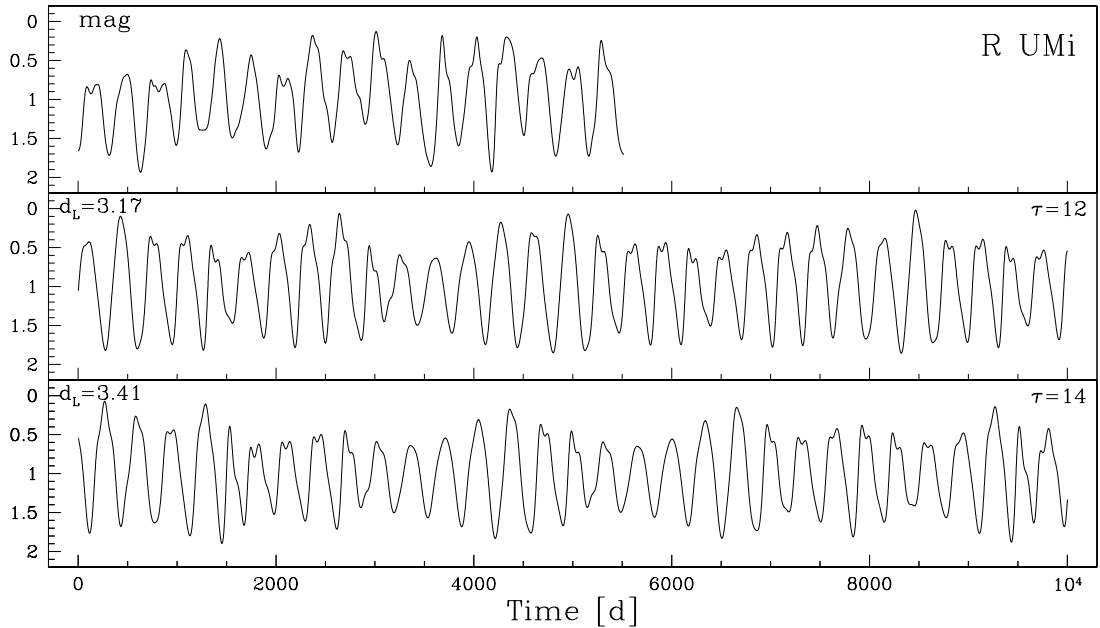}
  \caption{Flow reconstructions with light curve magnitude of R~UMi.}
   \label{rumis}
\end{figure}

We have also made flow reconstructions with the luminosities (obtained from
transforming the smoothed magnitude data) which are shown in
Fig.~\ref{sxhersL}.  This reconstruction is seen to lead to better synthetic
signals, although again only for two values of $\tau$ = 7 and 8.  However,
physically, the results are not that different.  One finds fractal
Lyapunov dimensions 3.2 -- 3.4, {\it i.e.} again larger than 3, but less than
4, and period ratios again close to and slightly greater than 2. (Here it has
not been possible to construct robust 5D maps).

\vskip 0.2cm

Although the quality of the reconstructions for SX~Her are hampered by the
shortness of the data set (not many modulation cycles), which shows up among
other things in a relatively poor stability of the maps and in a narrow range
of values of $\tau$ over which the reconstruction is possible, one can
reasonably conclude that that the embedding dimension is 4 and that the fractal
dimension is between 3.1 and $3.8 < 4$.

\subsection{The star R Ursae Minoris}

In Figure~\ref{rumis} on top we display the smoothed 5500\th d long R~UMi
observational signal that we have used to train our maps.  Underneath are two
segments of synthetic signals that were generated from the iteration of four
different reconstructed 4D maps, for values of $\tau$ going from 4 to 10
(days).  Good synthetic signals are seen to be obtainable from a range of
$\tau$ values, even though there are some intermediate values for which no good
reconstruction has been obtained.  The best 3D maps that we can construct are
incapable of producing synthetic signals that bear a resemblance to the
observational light curve, suggesting an embedding dimension $d_e=4$.

The maps for R~UMi are not as robust as those for R~Sct, even though the
observational accuracy is much higher.  One of the probable reasons is that
there are fewer cycles in the data.  Indeed, for a successful flow
reconstruction to be possible the data has to sample sufficiently well the
dominant features of the dynamics in phase space.  If some regions are only
covered very lightly, then the maps can have 'leaks' in such regions, and the
lack of robustness manifests itself in trajectories that blow up quickly.  The
synthetic signals that are constructed from the maps are therefore relatively
short.  It is also hard to obtain accurate Lyapunov exponents and fractal
dimensions.

As for AC~Her, the R~UMi signal does not explore very well the linear
neighborhood of the fixed point.  Consequently it is not possible to derive any
information about a possible resonance.

We find that the fractal dimensions for good synthetic signals hover around
$d_L= 3.25 \pm 0.15$.  Reassuringly this dimension is found to be stable in the
sense that reconstructions in 5D also yield $d_L$ in the same range, and
importantly that they are less than 4, {\it i.e.} independent of the dimension
of the embedding space.  The conclusion that imposes itself, at least
tentatively, is that the physical dimension which is sandwiched between $d_L
\sim 3.25$ and $d_e=4$, is $d=4$ for this star as well.

\begin{figure}
  \includegraphics[height=0.38\textheight]{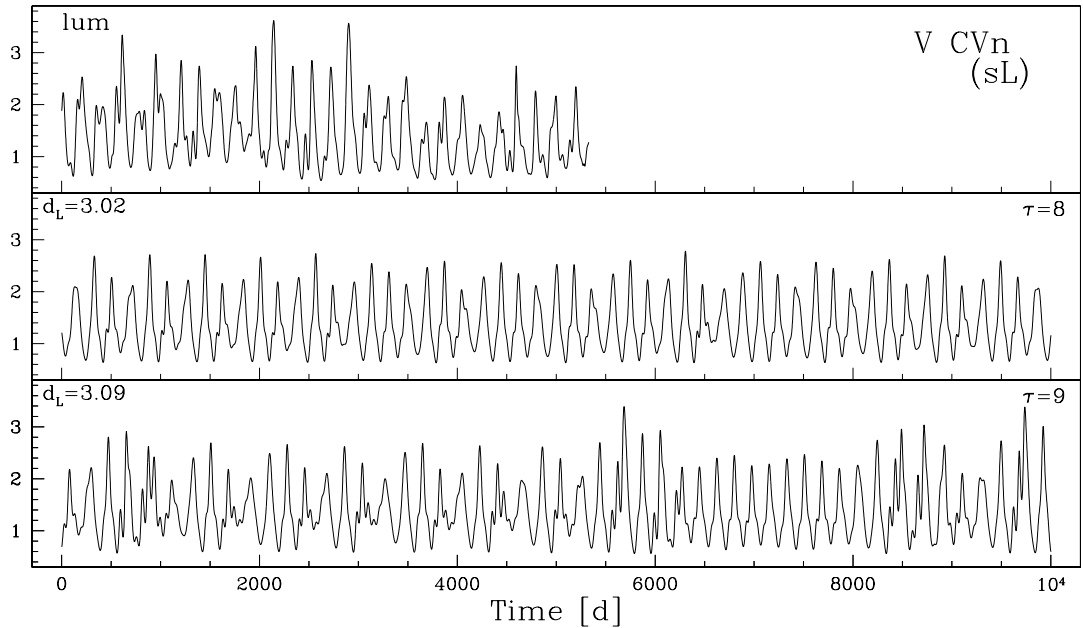}
  \caption{Flow reconstructions with the stellar  luminosity of V~CVn.}
  \label{vcvnsL}
\end{figure}

\subsection{The star V Canum Venaticorum}

We have not been able to make a reconstruction with the smoothed magnitude data
themselves, even with various values of smoothing, although the light curve
data span some 27 cycles.  The reason why our reconstruction is not successful
could be that over the 15 years of observations the light curve does not
explore its whole potential, {\it i.e.} it does not explore enough of the
physical phase space to allow us to make a successful reconstruction.  We
cannot exclude the possibility that a higher order dynamics might be at work,
but then again we would need more observational information to lay it bare.
Perhaps the star undergoes intermittent mass loss which causes obscuration and
leads to variations in the overall luminosity.  We think the first of these
former may be the problem, because, after a prior conversion of the same
magnitude data to luminosities we arrive at a reasonably good reconstruction
(with two values of $\tau$, 8 and 9).  The smoothed observational luminosity
data that are used in the reconstruction are shown in Fig.~\ref{vcvnsL}
together with two typical synthetic luminosity curves.

\begin{figure}
  \includegraphics[height=0.38\textheight]{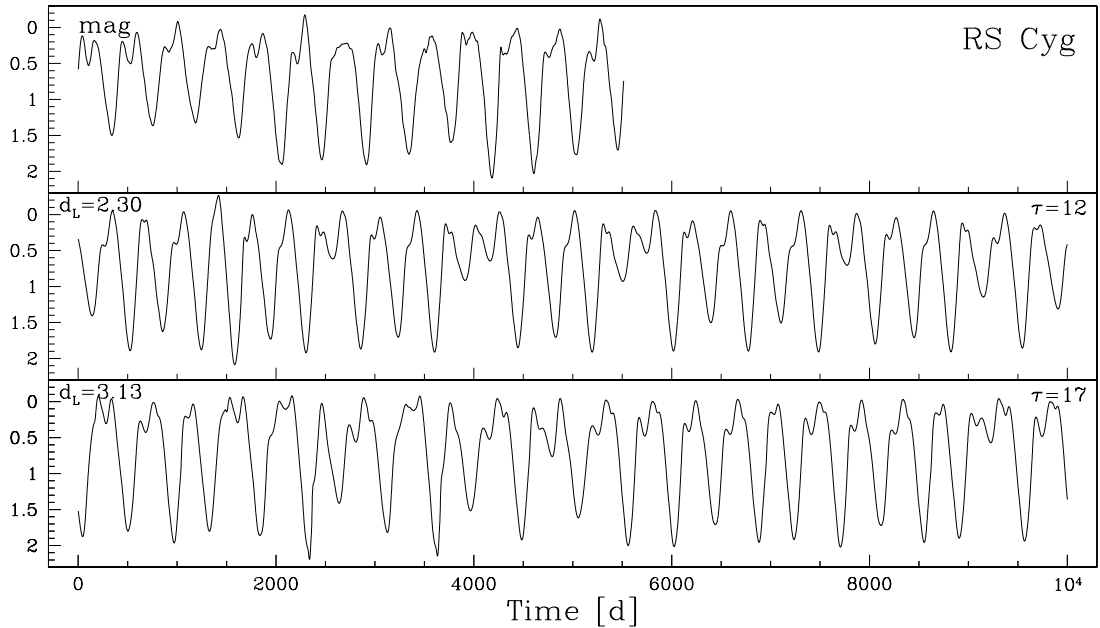}
  \caption{Flow reconstructions with light curve magnitude of RS~Cyg.}
  \label{rscygs}
\end{figure}

In conclusion, the data for V~CVn are just barely sufficient for a
reconstruction which therefore remains fragile.

\subsection{The star RS Cygni}

The smoothed observed light curve of RS~Cyg is displayed in Fig.~\ref{rscygs},
followed by a couple of 4D synthetic light curves.  The flow reconstructions
are reasonably robust (with a range of $\tau$ from 11 to 17).  The synthetic
light curves appear good, but have some difficulties capturing the moving small
RV~Tau-like feature, and they have added wiggles that are nonexistent in the
data.  These difficulties are not astonishing because of the shortness of the
data set, {\it viz.} only about 13 cycles -- we should not expect to get more
out the data than is in them.  A comparison of the amplitude Fourier spectrum
of the synthetic light curves with that of the data also shows that the
reconstruction is somewhat lacking.  The Lyapunov dimensions have a wide range
of values 2.1 -- 3.1.  A 2:1 resonance condition is approximately satisfied for
$\tau$ = 11, 12 and 15.

We find that flow reconstructions in 3D are not satisfactory.  As to 5D there
are too few data points to attempt a reconstruction, but our results, in
particular Fig.~\ref{rscygs}, suggest that for RS~Cyg an embedding dimension
$d_e=4$ is sufficient.  The physical phase space dimension $d$ is probably
equal to 4 in order to accommodate two complex vibrational modes (strictly
speaking, if $d_L < 3$ a value of 3 cannot be ruled out from $d_L <d \leq
d_e$).  Possibly there is again a 2:1 resonance condition between the two modes
in this star.  The analysis of RS~Cyg is thus promising, but for a reliable
flow reconstruction it will be necessary to gather more information about the
light curve of this star.

\section{Discussion and Conclusions}

The reader may have been left wondering why the classical Cepheids undergo
periodic pulsations while their neighbors in the Hertzsprung-Russell diagram
pulsate irregularly.  We want to address the difference between these two
classes of stars briefly.  They have the same luminosity range, but the
irregular stars have lower masses, by about a factor of 10.  As a result the
coupling between the heat flow and the acoustic oscillation is strongly
enhanced.  This can best be seen with a linear stability analysis of the
equilibrium stellar models.  Let us call $\omega$ the eigenvalues of the
pulsation modes (for an assumed ${\rm exp} (i \omega\thinspace t)$ dependence).
Then for the modes that are relevant here, {\it i.e.}  the lowest frequency
modes, we find that the relative growth rates of these modes $|{\rm
Im}\thinspace \omega /{\rm Re}\thinspace \omega|$ are of the order of a few
percent for the classical Cepheids, but they are of order unity for the
semiregular stars.  The mathematical consequence is that there exists a center
manifold for the classical Cepheids, and their behavior can be captured with
amplitude equations (normal forms) \cite{b93}; in particular, the periodic
pulsations of the classical Cepheids owe their existence to the proximity of a
broad Hopf bifurcation.

Clearly, a large value of the relative growth rates is a prerequisite for the
occurrence of chaos (appreciable amplitude changes must be possible over a
period).  For the W~Virginis type stars at the low luminosity end the growth
rates are still small, and periodic pulsations (limit cycles) prevail, just as
for the classical Cepheids.  For the more luminous ones, such as the stars
analyzed in this paper, the growth rates are of order unity, thus fulfilling
the necessary condition for chaotic behavior.

We have applied the global flow reconstruction technique to the high quality
observational data of SX~Her, R~UMi, V~CVn and RS~Cyg.  In view of the
relatively small number of observed cycles and the complexity of the light
curves our conclusions must be taken with some caution.  The flow
reconstructions lack the robustness of those obtained earlier from the large
amplitude variable star R~Sct.  Consequently, it is not possible to produce
very long synthetic light curves by iterating the map, and the Lyapunov
exponents and the fractal dimension are therefore beset with large fluctuations
from one synthetic signal to another.  Our flow reconstructions are best for
SX~Her and R~UMi.  For RS~Cyg the time span of the observed light curve is
clearly too short.  Our poor success with V~CVn is perhaps related to the
complexity of its light curve.

We find that the light curves of six different large amplitude, irregularly
pulsating stars all indicate a minimum embedding dimension of 4, although the
only reconstructions that have some robustness are those for R~Sct and SX~Her.
For both of these the fractal Lyapunov dimensions fall in the range 3.1 -- 3.5,
and one can conclude that the dimension of the physical phase space of the
dynamics is 4.  This in turn suggests that the 'natural' generalized
coordinates in this phase space are the complex amplitudes of two vibrational
modes.

Our analysis of both R~Sct and SX~Her suggests that the irregular behavior has
the same physical mechanism.  The irregular behavior arises through the
nonlinear interaction between two vibrational modes, one of lower frequency
that is unstable and one with higher frequency that is stable.  Furthermore,
the ratio of the two frequencies is close to 2:1 in both stars.  We thus arrive
at the same physical picture of the underlying mechanism for the irregular
light curve.  A lower frequency mode is self-excited and entrains an otherwise
linearly stable overtone through a low order (2:1) resonance.  The irregular
pulsation occurs as a result of continual exchange of energy between the two
resonant modes.  The minimum embedding dimensions of the other four stars
corroborate this finding although no information about a resonance could be
gotten from the linearization of the map.

The fact that the underlying dynamics of these types of stars are low
dimensional and chaotic is not surprising, and it was in fact predicted by
numerical hydrodynamical simulations of W~Vir type stars \cite{serre96}
\cite{wvir}.  Furthermore a topological analysis of the attractor
\cite{letellier} corroborated these conclusions.

The flow reconstruction technique thus has shed new light on the old mystery of
the nature of the irregular pulsations of a large class of large amplitude
pulsating variable stars..

\begin{theacknowledgments}
Work supported by the National Science Foundation and by MTA-OTKA (T-026031).
\end{theacknowledgments}


\doingARLO[\bibliographystyle{aipproc}]
          {\ifthenelse{\equal{\AIPcitestyleselect}{num}}
             {\bibliographystyle{arlonum}}
             {\bibliographystyle{arlobib}}
          }


\end{document}